\documentclass[10pt,twocolumn,letterpaper]{article}

\usepackage[pagenumbers]{wacv} 

\usepackage{graphicx}
\usepackage{amsmath}
\usepackage{amssymb}
\usepackage{booktabs}
\usepackage{multirow} 
\usepackage{flushend}

\usepackage[pagebackref,breaklinks,colorlinks]{hyperref}

\usepackage[capitalize]{cleveref}
\crefname{section}{Sec.}{Secs.}
\Crefname{section}{Section}{Sections}
\Crefname{table}{Table}{Tables}
\crefname{table}{Tab.}{Tabs.}

\def\wacvPaperID{*****} 
\def\confName{WACV}
\def\confYear{2025}

\begin{document}

\title{Sparse Mixture-of-Experts for Non-Uniform Noise Reduction in MRI Images}

\author{Zeyun Deng \hspace{2em} Joseph Campbell\\
Purdue University\\
{\tt\small \{deng334, joecamp\}@purdue.edu}
}

\maketitle

\begin{abstract}
Magnetic Resonance Imaging (MRI) is an essential diagnostic tool in clinical settings, but its utility is often hindered by noise artifacts introduced during the imaging process. Effective denoising is critical for enhancing image quality while preserving anatomical structures. However, traditional denoising methods, which often assume uniform noise distributions, struggle to handle the non-uniform noise commonly present in MRI images. 
Building on prior multi-branch MRI denoising approaches, we introduce a fine-grained sparse mixture-of-experts framework for MRI image denoising.
Our method decomposes each image into patch-based or segmentation-based regions, groups regions according to their learned feature similarity, and routes each region to a specialized denoising convolutional neural network.
Our method demonstrates superior performance over state-of-the-art denoising techniques on both synthetic and real-world brain MRI datasets. Furthermore, we show that it generalizes effectively to unseen datasets, highlighting its robustness and adaptability.
\end{abstract}

\section{Introduction}
Magnetic Resonance Imaging (MRI) is a vital imaging technique in clinical diagnostics, offering detailed insights into the brain's internal structures. However, MRI is prone to various forms of noise that can obscure critical information, affecting the quality and accuracy of diagnosis~\cite{macovski1996noise}.
In many vision applications, image noise is commonly assumed to be Gaussian and zero-mean~\cite{buades2005non, zhang2017beyond}.
However, the unique acquisition process associated with MRI results in a spatially-varying noise signal which follows a non-zero-mean distribution~\cite{gudbjartsson1995rician}.
This is due to many factors, including magnetic field inhomogeneity, coil non-uniformity, heterogeneous tissue properties, and patient motion.
At the same time, MRI images have a fixed underlying structure, as anatomical features of the brain remain relatively consistent across individuals.
The result is that different regions within the brain may exhibit distinct noise patterns, influenced by tissue composition and scanner settings~\cite{manson2023evaluation}.
In this paper, we develop methods for removing non-uniform, spatially-varying noise to generate clean, high-quality brain MRI images for reliable clinical interpretation.

While general image denoising methods are designed to handle a wide range of images~\cite{zhang2017beyond}, our setting allows us to exploit the stable anatomical structure of the brain to better address spatially varying noise. HydraNet~\cite{gregory2021hydranet} introduces a multi-branch, patch-based framework that routes patches to specialized denoisers using noise-level or location-based assignment strategies. Although effective, these predefined categories and coarse spatial assignments may not capture finer-grained structural and noise variations, motivating our more granular cluster-based routing approach.

To address this shortcoming, we propose a more granular sparse mixture-of-experts (MoE) approach to MRI denoising.
Our method uses image segmentation techniques~\cite{kirillov2023segment, ma2024segment} to decompose an image into fine-grained structures, with the goal of independently denoising each structure according to its unique noise characteristics.
Each expert in the mixture is a specialized denoising model, fine-tuned to remove noise specific to its assigned structures.

During training, segmented structures are clustered based on their similarity, and a dedicated expert is fine-tuned on each cluster, ensuring that the model adapts to the unique noise properties of each group.
At test time, segmented structures are mapped to their corresponding expert for denoising, and the denoised structures are seamlessly reconstructed into a single high-quality image.
This approach ensures region-specific noise reduction while preserving anatomical details across the entire image.

The main contributions of this paper are as follows:
\begin{itemize}
    \itemsep-0.2em 
    \item \textbf{Sparse MoE Denoising:} We propose a sparse mixture-of-experts approach to MRI denoising and empirically show it out-performs state-of-the-art baselines over both synthetic and real-world datasets.
    
    \item \textbf{Granularity Evaluation:} We quantitatively and qualitatively evaluate three variants of our approach with different levels of structural granularity and show that higher granularity tends to produce better results.
    
    \item \textbf{Generalizability and Robustness Evaluation:} We evaluate our approach on novel datasets and show that it generalizes well to unseen images and noise levels.
\end{itemize}

\section{Related Works}

Denoising in MRI has a long history of development, spanning from traditional filtering techniques to advanced deep learning methods. Basic filtering methods, such as Gaussian and median filtering, have been widely used for noise reduction ~\cite{gonzalez2018digital}. These approaches work by smoothing image regions to suppress noise but often lead to a loss of fine structural details, which is undesirable for clinical MRI. Wavelet-based filtering methods ~\cite{ouahabi2013review} and non-local means (NLM) filtering ~\cite{buades2005non} offer improved preservation of edges and details but can struggle with complex noise patterns and structural variability across MRI scans.

With the advent of deep learning, Convolutional Neural Networks (CNNs) have become a popular choice for denoising due to their ability to learn complex noise characteristics directly from data. Several CNN architectures have been proposed for image denoising, each offering unique strategies to improve noise reduction and structural preservation in MRI images. Among them, the Denoising Convolutional Neural Network (DnCNN)~\cite{zhang2017beyond}, Shrinkage Convolutional Neural Network (SCNN) ~\cite{isogawa2017deep}, and Deep Learning-based Reduction (dDLR) ~\cite{kidoh2020deep, tajima2023usefulness} are notable for their distinct approaches to handling noise in medical images. 

DnCNN, the foundation of our proposed method, uses a residual learning strategy that models the noise component directly, enabling effective removal of Gaussian and mixed noise by subtracting the learned noise from the original image. SCNN, on the other hand, introduces a soft shrinkage mechanism applied to feature maps, selectively emphasizing relevant features while suppressing noise. Meanwhile, dDLR combines CNN-based denoising with a unique high-frequency-focused regularization approach using discrete cosine transform (DCT). This ensures noise suppression while maintaining anatomical details. Together, these methods showcase the evolution of deep learning-based MRI denoising, moving from general noise reduction to approaches that integrate domain-specific insights for improved structural consistency and detail preservation. 

Patch-based methods have advanced significantly in recent years. The Fuzzy Gaussian Mixture Model (FGMM) clustering method~\cite{shi2021mr} builds on the Gaussian Mixture Model (GMM) by introducing fuzziness into the clustering process, enabling better handling of overlapping image patches. While FGMM optimizes a cost function using the Half Quadratic Splitting method during training and achieves good denoising performance, its reliance on large-scale patch sampling and computationally expensive iterative optimization may introduce challenges for scalability and broader applicability in clinical scenarios.

Similarly, geometric structure clustering methods~\cite{fei2013patch} use patch-wise principal component analysis (PCA) and clustering based on dominant geometric orientations to group patches with similar structures. This approach applies hard thresholding in the PCA domain to selectively remove noise while preserving critical geometric details such as edges and textures. However, the focus on dominant orientations may lead to challenges in capturing intricate anatomical structures and intensity variations present in medical images, particularly in highly heterogeneous regions.

Although both methods provide valuable insights, their assumptions and limitations may make them less effective for the specific challenges of denoising MRI images with spatially varying noise, which our proposed region-specific denoising strategies aim to address. By leveraging clustering and segmentation tailored to MRI's anatomical structures, our methods offer enhanced adaptability and performance for such complex noise scenarios.

Encoder-decoder architectures have also been explored in denoising ~\cite{ronneberger2015u, casas2021adversarial}. Encoder-decoder models extract low-dimensional features, often combining them with upsampling layers to reconstruct denoised images, leveraging feature abstraction for improved denoising and robustness.  

HydraNet~\cite{gregory2021hydranet} introduces a multi-branch, patch-based framework for MRI denoising with spatially varying noise. It uses separate DnCNN denoisers with assignment based on PSNR, residual standard deviation, or sagittal slice location. The first two variants use predefined noise categories with SSIM-based matching, while the location variant routes patches to left, middle, or right brain denoisers. These assignments may miss finer-grained structural and noise variations. Our method instead uses feature-based clustering with patch-based or segmentation-based decomposition for more granular expert assignment.

\begin{figure*}[t]
    \centering
\includegraphics[width=\textwidth]{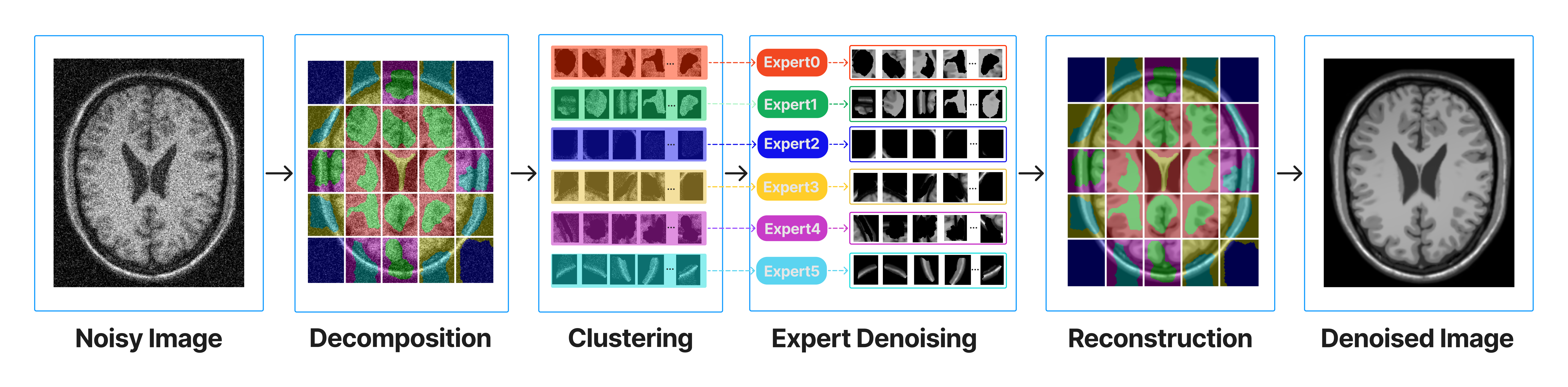}
 \vspace{-2.5em}
    \caption{Our proposed sparse mixture-of-experts model for region-specific denoising consists of the following steps. \textbf{Decomposition:} The noisy input image is decomposed into fine-grained regions. \textbf{Clustering and Expert Denoising:} Each region is mapped to a cluster according to its noise profile and denoised with the corresponding expert denoising model. \textbf{Reconstruction:} The denoised regions are then recombined to obtain the final high-quality denoised image. \textit{Note: Image noise is exaggerated for visualization purposes.}}
    \label{fig:overview}
\end{figure*}

\section{Methodology}
\label{sec:methodology}

We follow a residual learning approach to denoising~\cite{zhang2017beyond}, in which we define a noisy image \(\mathbf{y}\) as the combination of a clean image \(\mathbf{x}\) and additive non-uniform noise \(\mathbf{v}\), such that:
\begin{equation}
\mathbf{y} = \mathbf{x} + \mathbf{v}.
\label{eq:noisy_image}
\end{equation}
Our goal is to learn a mapping function \(f\) such that $f(\mathbf{y}) \approx \mathbf{v}$, enabling recovery of the clean image as:
\begin{equation}
\mathbf{\hat{x}} = \mathbf{y} - f(\mathbf{y}).
\label{eq:clean_image}
\end{equation}

In the presence of spatially non-uniform noise, learning a single mapping function \(f\) is inherently challenging as the additive noise $\textbf{v}$ is often assumed to be zero-mean (and usually Gaussian) with respect to the signal~\cite{buades2005non, zhang2017beyond}.
However, in MRI non-uniformity means that the noise follows a non-zero-mean Rician distribution which is signal-dependent and thus spatially-varying.
Consequently, region-specific denoising strategies become essential to address these variations, as they enable models to adapt to local noise patterns and better preserve structural details.
This limitation of global models highlights the need for approaches that incorporate clustering or segmentation to tailor denoisers to the unique characteristics of each region.

To address this, we decompose an image \(\mathbf{y}\) into \(M\) regions \(\{\mathbf{y}_1, \mathbf{y}_2, \dots, \mathbf{y}_M\}\), each exhibiting unique noise characteristics.
We employ \(K\) experts \(f_1, f_2, \dots, f_K\), with each expert specializing in denoising regions sharing similar noise characteristics.
A gating mechanism \(g(\mathbf{y}_j)\) determines the appropriate expert for a given region \(\mathbf{y}_j\). Each region is assigned to one of \(K\) clusters, denoted as \(C_j \in \{1, 2, \dots, K\}\), and the gating function outputs a one-hot vector of length \(K\): 

\begin{equation}
g(\mathbf{y}_j) = [g_1(\mathbf{y}_j), g_2(\mathbf{y}_j), \dots, g_K(\mathbf{y}_j)],
\label{eq:gating_function}
\end{equation}
where:
\begin{equation}
g_i(\mathbf{y}_j) =
\begin{cases}
1, & \text{if } i = C_j, \\
0, & \text{otherwise}.
\end{cases}
\label{eq:gating_mechanism}
\end{equation}

The gating mechanism activates the expert corresponding to the assigned cluster \(C_j\), ensuring that each region is processed by the most suitable denoising model.
We refer to this formulation as a sparse mixture-of-experts since only a single expert is activated at a time -- as determined by the gating function.
This approach ensures each region is processed by the expert best suited to its noise profile, enabling effective denoising of spatially varying noise.

Formally, the noise estimate for each region is given by:
\begin{equation}
\mathbf{v}_j = \sum_{i=1}^{N} g_i(\mathbf{y}_j) f_i(\mathbf{y}_j).
\label{eq:noise_estimate}
\end{equation}
Finally, the denoised image regions \(\mathbf{x}_j\) are obtained with:
\begin{equation}
\mathbf{\hat{x}}_j = \mathbf{y}_j - \mathbf{v}_j.
\label{eq:clean_region}
\end{equation}
The full denoised image is then obtained by recombining all denoised regions:
\begin{equation}
\mathbf{\hat{x}} = \text{Merge}(\mathbf{\hat{x}}_1, \mathbf{\hat{x}}_2, \dots, \mathbf{\hat{x}}_M),
\label{eq:reconstructed_image}
\end{equation}
where overlapping areas are averaged to ensure smooth transitions and avoid undesirable artifacts.

This formulation leverages the mixture-of-experts model to capture the region-specific noise characteristics, enabling precise denoising tailored to each part of the image. An overview of the proposed framework is shown in \Cref{fig:overview}.

\subsection{Image Decomposition}
Decomposing the noisy image \(\mathbf{y}\) into smaller, manageable regions is a critical step to account for spatially varying noise characteristics. This decomposition facilitates tailored denoising strategies for different parts of the image.
We discuss two approaches -- patch-based and segmentation-based decomposition -- to better understand the relationship between denoising performance and region granularity.
The patch-based approach results in a simpler, coarser decomposition while the segmentation-based approach attempts to identify granular anatomical structures within the MRI image.

\subsubsection{Patch-Based Decomposition}
Our patch-based decomposition method applies a straightforward sub-division of the image \(\mathbf{y}\) into uniform regions of size \(48 \times 48\).
We induce overlap among the regions by using a stride of 20 when generating patches.
This is to avoid discontinuous ``jumps'' in pixel values when recombining image regions in a later step, as such jumps produce undesirable artifacts in the final denoised image.
Overlap allows for pixel averaging and smooth transitions between regions.

\subsubsection{Segmentation-Based Decomposition}
Our segmentation-based method further decomposes each patch obtained above into fine-grained anatomical regions defined by variations in their pixel-wise intensity and structural boundaries.
We apply a pre-trained segmentation model to each patch to extract segmentation masks for regions of interest.
From each mask we create two regions: one containing the area within the mask, with all pixels outside the mask set to zero; and the inverse containing the area outside the mask with all pixels inside set to zero so as to preserve the background.

In this work we evaluate two segmentation models: Segment Anything Model (SAM)~\cite{kirillov2023segment} and a variant designed to segment medical images, MedSAM~\cite{ma2024segment}.
While SAM is a general-purpose segmentation model, MedSAM is specifically fine-tuned for medical imaging tasks and is expected to better capture domain-specific nuances in MRI images.

\subsection{Cluster-Based Gating}
After dividing our input image into patches using one of the above decomposition methods, we group the patches according to their image similarity.
We assume that similar regions contain similar noise characteristics, and therefore similar regions should be attended to by the same expert denoising model.
We first extract latent feature embeddings from each region using a pre-trained ResNet18 model~\cite{he2016deep}.
From these embeddings, we extract the most meaningful features by applying dimensionality reduction via principal component analysis which yields a final feature vector of size $256$ for each region.
We group the regions according to their feature similarity by applying the k-means clustering algorithm over the set of feature vectors.
The clustering result informs the gating function in \cref{eq:gating_function}, which determines which expert \(f_i\) should denoise a specific region \(\mathbf{y}_j\).
Mathematically, this is achieved by assigning each region \(\mathbf{y}_j\) to one of \(k\) clusters as obtained with k-means, with the cluster assignment denoted as \(C_j \in \{1, 2, \dots, k\}\).

\subsection{Learning a Residual Mapping}
Each expert \(f_i(\mathbf{y}_j)\) specializes in learning the noise pattern for a specific cluster. The noise estimate for a region \(\mathbf{v}_j\) is computed as described in \cref{eq:noise_estimate}, using the gating mechanism and expert outputs. The clean region \(\mathbf{\hat{x}}_j\) is then reconstructed as defined in \cref{eq:clean_region}, by subtracting the noise estimate from the noisy input.

Each expert is a DnCNN model~\cite{zhang2017beyond} which has been fine-tuned over the regions assigned to the expert's cluster.
When fine-tuning, we start with a pre-trained model and freeze all layers except for the final layer to avoid overfitting. The DnCNN architecture includes an initial convolutional layer with 64 kernels of size \(3 \times 3\), followed by a ReLU activation. This is followed by 15 convolutional layers, each with 64 kernels of size \(3 \times 3\), batch normalization, and ReLU activation. A final output layer with a single \(3 \times 3\) convolutional kernel produces the residual image. All convolutional layers use a stride of \((1, 1)\) and a padding of \((1, 1)\) to ensure that the output image is the same size as the input image.

\subsection{Reconstruction and Merging}

The final denoised image \(\mathbf{\hat{x}}\) is reconstructed by recombining all denoised regions as described in \cref{eq:reconstructed_image}. Overlapping areas are averaged during the merging process to ensure seamless transitions between adjacent regions.

\section{Experiments}

We design experiments to answer the following questions.
\textbf{1)} Does our proposed approach result in better denoising performance than state-of-the-art baselines?
\textbf{2)} Can our approach generalize to novel MRI images, which may contain different noise patterns if they are collected from different machines and patients?
\textbf{3)} What is the relationship between denoising performance and region granularity?

\subsection{Datasets}

We evaluate our approach over two datasets: BrainWeb, a synthetic dataset, and IXI, a real-world dataset.
The combination of these datasets provides a balance between controlled noise levels for benchmarking and realistic noise conditions for real-world validation, ensuring a robust evaluation of the proposed denoising methods under diverse conditions.

\begin{itemize}
    \item \textbf{BrainWeb~\cite{cocosco1997online}:} The BrainWeb dataset, a widely used synthetic MRI dataset, offers controlled scenarios with varying noise levels and intensity non-uniformities (RF inhomogeneities). Key features include noise levels of 0\%, 1\%, 3\%, and 5\%, along with RF inhomogeneity levels of 0\%, 20\%, and 40\%, simulating intensity non-uniformities. The dataset is divided as follows: 
    the training set contains 36 images across all noise and RF levels, the validation set includes 12 images, and the test set consists of 57 images.

    \item \textbf{IXI~\cite{ixi_dataset}:} The IXI dataset was preprocessed using the registration procedure described in \cite{chen2022transmorph} to ensure spatial alignment and consistency across images. To simulate varying noise conditions, Rician noise with standard deviations ranging from 0\% to 5\% was artificially added. The dataset is divided as follows:
    the training set consists of 40 images with varying noise levels, the validation set contains 10 images, and the test set includes 130 images, covering a broad range of anatomical features and noise characteristics.
\end{itemize}

\begin{table*}[t]
\centering
\parbox{.48\linewidth}{
\caption{Denoising results for each method when trained over \textbf{BrainWeb}. Left columns show results over a held-out test set and right columns show results over a completely novel dataset (IXI).}
\label{tab:brainweb_results}
\resizebox{\columnwidth}{!}{%
\begin{tabular}{@{}lcccc@{}}
\toprule
\multirow{2}{*}{\textbf{Model}} & \multicolumn{2}{c}{\textbf{BrainWeb Test Data}} & \multicolumn{2}{c}{\textbf{IXI Test Data}} \\ \cmidrule(l){2-5} 
                                & \textbf{SSIM}    & \textbf{PSNR (dB)}          & \textbf{SSIM}   & \textbf{PSNR (dB)}         \\ \midrule
DnCNN                           & 0.8435           & 34.51                       & 0.8143          & 39.90                      \\
DnCNN (Fine-tuned)              & 0.8478           & 33.55                       & 0.8082          & 38.22                      \\
HydraNet                        & 0.8874           & 38.76                       & 0.8392          & 41.36                      \\
\midrule
SMoE Patch                      & \textbf{0.9318}  & \textbf{39.65}              & 0.8709          & 40.56                      \\
SMoE Seg. (SAM)              & 0.9232           & 39.57                       & 0.9090          & 43.39                      \\
SMoE Seg. (MedSAM)           & 0.9245           & 39.44                       & \textbf{0.9217} & \textbf{43.49}             \\ \bottomrule
\end{tabular}%
}
}
\hfill
\parbox{.48\linewidth}{
\caption{Denoising results for each method when trained over \textbf{IXI}. Left columns show results over a held-out test set and right columns show results over a completely novel dataset (BrainWeb).}
\label{tab:ixi_results}
\resizebox{\columnwidth}{!}{%
\begin{tabular}{@{}lcccc@{}}
\toprule
\multirow{2}{*}{\textbf{Model}} & \multicolumn{2}{c}{\textbf{IXI Test Data}}       & \multicolumn{2}{c}{\textbf{BrainWeb Test Data}} \\ \cmidrule(l){2-5} 
                                & \textbf{SSIM}    & \textbf{PSNR (dB)}          & \textbf{SSIM}   & \textbf{PSNR (dB)}         \\ \midrule
DnCNN                           & 0.8143          & 39.90                      & 0.8435         & 34.51                      \\
DnCNN (Fine-tuned)              & 0.8022          & 38.42                      & 0.8381         & 33.27                      \\
HydraNet                        & 0.9144          & 44.91                      & 0.8845         & 35.34                      \\
\midrule
SMoE Patch                      & \textbf{0.9383} & \textbf{46.60}             & \textbf{0.9025} & 37.19                      \\
SMoE Seg. (SAM)              & 0.9230          & 45.57                      & 0.8889         & \textbf{37.81}             \\
SMoE Seg. (MedSAM)           & 0.9318          & 45.81                      & 0.8898         & 37.58                      \\ \bottomrule
\end{tabular}%
}
}
\end{table*}

\begin{figure*}[t]
    \centering
    \includegraphics[trim={3cm 0 3cm 3cm},clip,width=\textwidth]{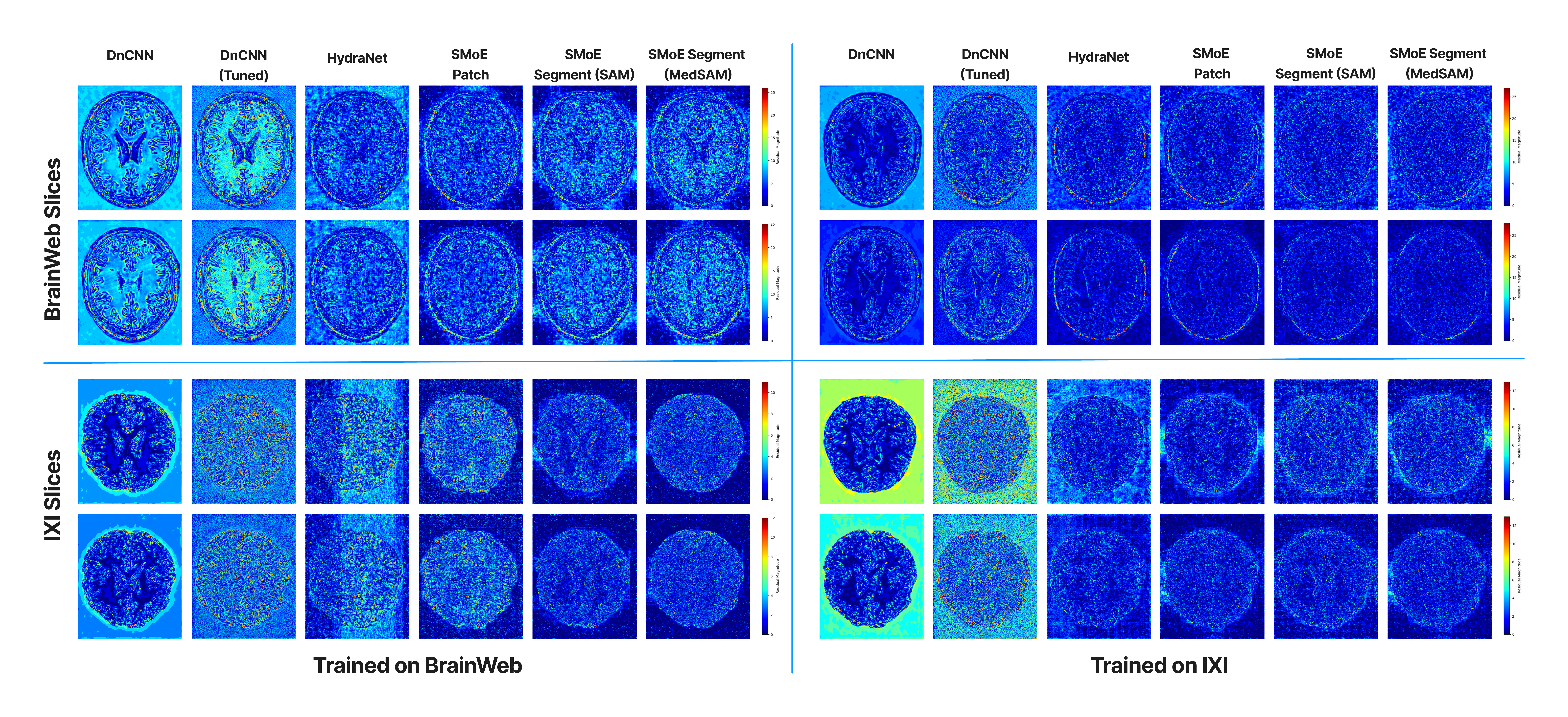}
    \vspace{-2.5em}
    \caption{Each figure shows the pixel-wise error between the denoised image and the ground truth image. Smaller (bluer) values are better.}
    \label{fig:residual}
\end{figure*}

\subsection{Evaluated Methods}

We evaluate three variants of our method, each corresponding to a different decomposition strategy from \cref{sec:methodology}, in addition to three state-of-the-art denoising methods.

\begin{itemize}
    \itemsep-0.2em
    
    \item \textbf{SMoE Patch:} Sparse mixture-of-experts with patch-based decomposition.

    \item \textbf{SMoE Segment (SAM):} Sparse mixture-of-experts with segmentation-based decomposition via SAM.

    \item \textbf{SMoE Segment (MedSAM):} Sparse mixture-of-experts with segmentation-based decomposition via MedSAM.

    \item \textbf{DnCNN~\cite{zhang2017beyond}:} Pre-trained DnCNN model with no fine-tuning.

    \item \textbf{DnCNN~\cite{zhang2017beyond} (Fine-tuned):} Pre-trained DnCNN model fine-tuned over each training dataset.

    \item \textbf{HydraNet~\cite{gregory2021hydranet}:}
A multi-branch model that routes patches by left, middle, or right brain location. We use this variant because it reports the highest SSIM and second-highest PSNR among the HydraNet variants.
\end{itemize}

\subsection{Evaluation Metrics}

We evaluate our SMoE model and the baselines both quantitatively and qualitatively. Quantitatively, we use peak signal-to-noise ratio (PSNR~\cite{hore2010image}) and structural similarity index measure (SSIM~\cite{wang2004image}), which together balance noise reduction and structural fidelity. PSNR quantifies the ratio of signal power to noise power, measuring denoising effectiveness, while SSIM evaluates structural preservation by comparing image similarity to ground truth, crucial for medical imaging.
Qualitatively, we compare pixel-wise error maps from denoised images to ground truth images to visually assess model performance.

\section{Results and Discussion}

\subsection{Denoising and Generalization Performance}

\begin{figure*}[t]
    \centering
    \includegraphics[trim={3cm 0 3cm 3cm},clip,width=\textwidth]{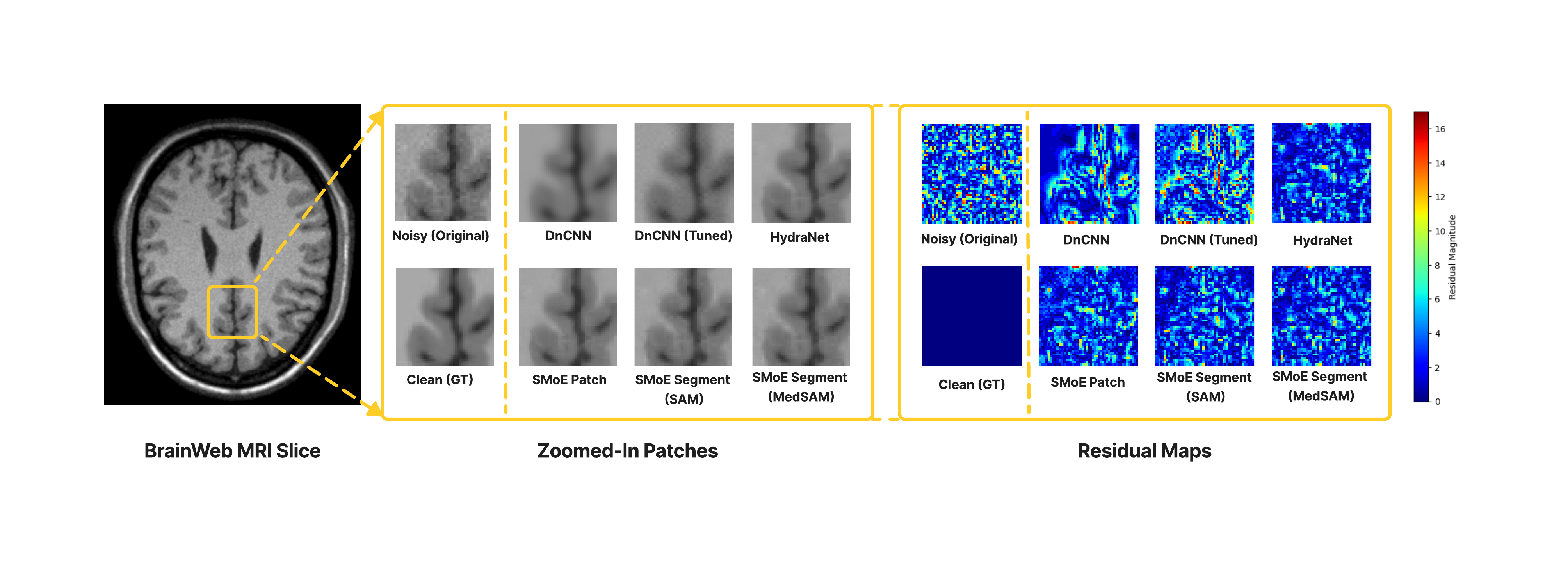}
    \vspace{-4.5em}
    \caption{Magnified denoised regions produced by each method and their corresponding pixel-wise error maps for a BrainWeb image.}
    \label{fig:brainweb_zoom}
\end{figure*}

\begin{figure*}[t]
    \centering
    \includegraphics[trim={3cm 0cm 3cm 3cm},clip,width=\textwidth]{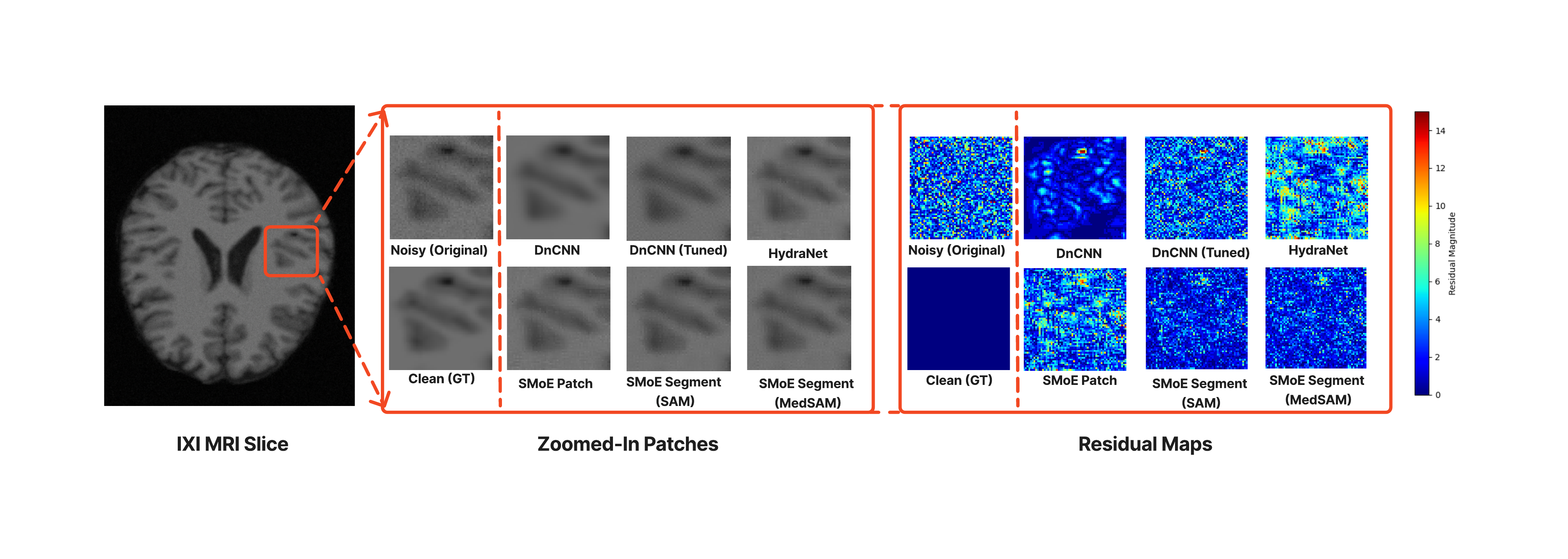}
    \vspace{-4.5em}
    \caption{Magnified denoised regions produced by each method and their corresponding pixel-wise error maps for an IXI image.}
    \label{fig:ixi_zoom}
\end{figure*}

The SSIM and PSNR values for each method and each dataset are given in \cref{tab:brainweb_results} and \cref{tab:ixi_results}.
From these results, we observe that \textbf{our proposed mixture-of-experts methods out-perform all baselines on both held-out test sets as well as completely novel datasets.}
Interestingly, quantitatively-speaking the patch-based decomposition approach (SMoE Patch) tends to out-perform our other decomposition methods.
However, as we see in the examples in \cref{fig:residual}, this slight advantage in SSIM and PSNR does not translate to a meaningful qualitative improvement and all of our proposed approaches denoise similarly well.
The one exception to this is when models are trained on BrainWeb and tested on IXI.
In this case, the segmentation-based methods produce much better results both quantitatively (\cref{tab:brainweb_results}, right) and qualitatively (\cref{fig:residual}, lower-left).
Zoomed-in patches and corresponding residual maps for BrainWeb and IXI are shown in \Cref{fig:brainweb_zoom} and \Cref{fig:ixi_zoom}, respectively.
For BrainWeb, DnCNN maintained structural shapes in the residual maps, indicating a strong blurring effect as it incorporates the underlying structural shape into its denoising process.
This behavior is further evident in the original zoomed-in patches, where the residuals reveal the impact of the blurring.
For IXI, segmentation-based methods produced notably smoother residuals, underscoring their robustness and ability to effectively suppress noise while preserving finer details.

\subsection{Generalization to Unseen Noise Levels}

In the real-world, images acquired from different patients, machines, and parameter configurations will result in different noise patterns and magnitudes.
Therefore, we wish for our denoising model to generalize to novel noise levels which may have been unseen in training.
To evaluate this, we trained each method over images from the BrainWeb dataset exhibiting noise at four different levels: 0\%, 1\%, 3\%, and 5\%.
We then denoised images from BrainWeb at two unseen noise levels: 7\% and 9\%.
From the results in \cref{tab:brainweb_unseen_results} we observe that all methods experience a significant degradation in both SSIM and PSNR at the 7\% noise level, and an even larger degradation at 9\%.
While our method exhibits the best performance with respect to PSNR, it falls behind in SSIM, particularly at the 9\% noise level.

This divergence between SSIM and PSNR highlights a trade-off in denoising: SSIM prioritizes structural fidelity, while PSNR emphasizes noise reduction~\cite{sara2019image}.
The choice depends on the task -- clinical applications may favor SSIM for structural accuracy, whereas quantitative analyses may benefit from higher PSNR.
The strong performance of segmentation-based methods at the 7\% level suggests higher region granularity helps balance these aspects.

\begin{table}[t]
\centering
\caption{Denoising results for each method when trained over BrainWeb with noise levels varying from 1\% to 5\% and tested over 7\% and 9\% noise levels.}
\label{tab:brainweb_unseen_results}
\resizebox{\columnwidth}{!}{%
\begin{tabular}{@{}lcccc@{}}
\toprule
\multirow{2}{*}{\textbf{Model}} & \multicolumn{2}{c}{\textbf{BrainWeb7}}       & \multicolumn{2}{c}{\textbf{BrainWeb9}}       \\ \cmidrule(l){2-5} 
                                & \textbf{SSIM}   & \textbf{PSNR (dB)}        & \textbf{SSIM}   & \textbf{PSNR (dB)}        \\ \midrule
DnCNN                           & \textbf{0.7412} & 27.97                     & \textbf{0.7183} & 26.04                     \\
DnCNN (Fine-tuned)              & 0.7216          & 27.59                     & 0.6889          & 25.85                     \\
HydraNet                        & 0.7034          & 28.49                     & 0.6375          & 25.75                     \\
\midrule
SMoE Patch                      & 0.7262          & \textbf{29.93}            & 0.6388          & \textbf{26.65}            \\
SMoE Seg. (SAM)              & 0.7163          & 29.15                     & 0.6422          & 25.92                     \\
SMoE Seg. (MedSAM)           & 0.7051          & 28.81                     & 0.6391          & 26.02                     \\ \bottomrule
\end{tabular}%
}
\end{table}

\subsection{Analysis of Region Granularity}

\begin{figure}[t]
    \centering
    \includegraphics[trim={3cm 0 3cm 3cm},clip,width=\linewidth]{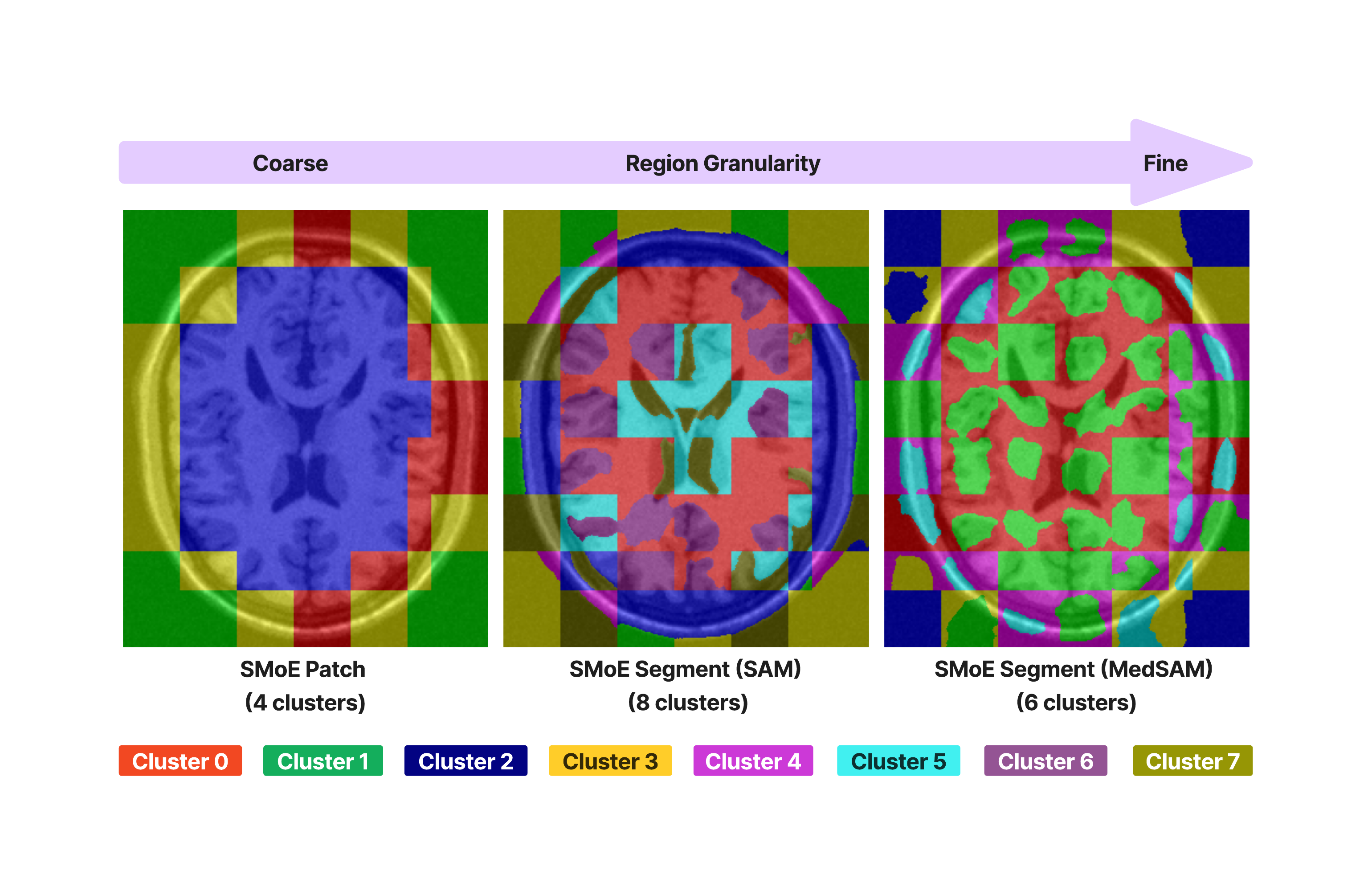}
    \caption{Visualization of the mapping between brain regions and clusters for each proposed decomposition method. Overlapping patch areas have been removed for visualization purposes.}
\label{fig:clustering_maps}
\end{figure}

We analyze the relationship between the granularity of regions resulting from image decomposition and corresponding denoising performance.
\Cref{fig:clustering_maps} visually illustrates the cluster assignments produced by the different segmentation approaches, highlighting how each method partitions the MRI images with varying granularity.
For the sake of visual clarity, the clustering maps are displayed without overlap.
Surprisingly, we note that in some cases MedSAM produces granular yet erroneous segmentations in the presence of noise, suggesting that fine-tuning SAM over medical images~\cite{ma2024segment} may have decreased overall robustness.

Due to this granularity, the optimal number of clusters varied across segmentation approaches, reflecting their ability to partition MRI images into meaningful regions. 
The SMoE Patch-based model resulted in 4 clusters, representing a relatively coarse division of image patches.
The SMoE Segment (MedSAM) model produced 6 clusters, offering more refined groupings by leveraging medical-specific segmentation masks.
In contrast, the SMoE Segment (SAM) model yielded 8 clusters, capturing finer structural details using generic segmentation masks.

We are interested in the following question: \textbf{is there a point at which increasing the granularity of the model results in decreased performance?}
The quantitative results in \cref{tab:brainweb_results} and \cref{tab:ixi_results} seem to suggest no; that increasing the granularity of decomposition does not meaningfully hurt performance, but may significantly improve it.
In particular, when models are trained on BrainWeb and tested on IXI (\cref{tab:brainweb_results}, right columns) there is a strong positive correlation between performance and granularity.
Qualitative results in \cref{fig:residual} and \cref{fig:ixi_zoom} support this result, as the segmentation-based methods yield improved denoising.

\begin{table}[t]
\centering
\caption{Gating function ablation for SMoE Patch where clusters are statically and randomly assigned. The first four rows show the values when each region is mapped to the same cluster, e.g. to Cluster 0. Random cluster indicates that all regions are assigned to a random cluster. Predicted cluster is our proposed method.}
\label{tab:patch_cluster_assignment}
\resizebox{0.65\columnwidth}{!}{%
\begin{tabular}{@{}lcc@{}}
\toprule
\textbf{Assigned Cluster} & \textbf{SSIM} & \textbf{PSNR (dB)} \\ \midrule
Cluster 0                 & 0.8625                & 38.59            \\
Cluster 1                 & 0.9060                & 34.81                      \\
Cluster 2                 & 0.9318                & 36.02                      \\
Cluster 3                 & 0.9141                & 37.98                      \\ \midrule
Random Cluster     & 0.9035                & 37.71     \\
Predicted Clusters & \textbf{0.9318}       & \textbf{39.65} \\ \bottomrule
\end{tabular}%
}
\end{table}

\subsection{Analysis of Expert Denoising Performance}

To better understand the impact of each expert on overall denoising performance, we performed ablation experiments in which the gating function in \cref{eq:gating_function} is removed.
We replaced it with two alternative gating functions: one in which every region is mapped to the same cluster, and another in which every region is mapped to a random cluster.
The results are shown in \cref{tab:patch_cluster_assignment} and show that \textbf{using the correctly predicted cluster results in better denoising performance than all ablated alternatives}.
This supports our hypothesis that the experts \textit{are} learning to reduce noise patterns present in specific regions of the MRI image.

There are several additional insights we can make with respect to the trade-off between noise reduction and structural preservation.
Cluster 0 achieved the highest PSNR (38.59 dB) but the lowest SSIM (0.8625), indicating effective global noise suppression at the cost of structural fidelity. This may be attributed to the fact that Cluster 0 corresponds to a small portion of the brain, as illustrated in \Cref{fig:clustering_maps}, focusing primarily on peripheral areas. As Cluster 0 contains a smaller and more homogeneous subset of the brain, it might be easier to train the expert denoiser for this cluster. The simpler task allows the denoiser to specialize in reducing noise effectively, leading to higher PSNR values.
In contrast, Cluster 2 yielded the highest SSIM (0.9318), highlighting superior structural preservation but with reduced noise suppression. Notably, Cluster 2 corresponds to the central area of the brain and covers the largest portion of each image, including critical and homogeneously represented anatomical structures. This extensive coverage likely contributes to its higher SSIM, as the central brain structures are pivotal to preserving structural fidelity.

\begin{table}[t]
\centering
\caption{Single-expert models trained over \textbf{BrainWeb} with noise levels varying from 1\% to 5\%. For comparison, $\star$ denotes the SMoE model with the best SSIM, while $\diamond$ denotes  the SMoE model with the best PSNR.}
\label{k1brainweb_results}
\resizebox{\columnwidth}{!}{%
\begin{tabular}{@{}lcccccc@{}}
\toprule
\textbf{Test Data}       & \textbf{Model}        & \textbf{SSIM}    & \textbf{PSNR (dB)} \\ \midrule
\multirow{3}{*}{BrainWeb} & Patch ($k=1$)               & 0.9019                   & 38.71                      \\
                         & Seg. (SAM, $k=1$)                 & \textbf{0.9327}                   & 39.57                      \\
                         & Seg. (MedSAM, $k=1$)              & 0.9186                   & 39.53                      \\ \cmidrule(lr){2-4}
                          & SMoE Patch$^{\star \diamond}$                     & 0.9318  & \textbf{39.65}                       \\ \midrule
\multirow{3}{*}{IXI}     & Patch ($k=1$)               & 0.8583                   & 42.05                      \\
                         & Seg. (SAM, $k=1$)                 & \textbf{0.9551}                   & \textbf{43.91}                     \\
                         & Seg. (MedSAM, $k=1$)              & 0.9331                   & 43.76                      \\ \cmidrule(lr){2-4}
                         & SMoE Seg. (MedSAM)$^{\star \diamond}$   & 0.9217 & 43.49  \\ \midrule
\multirow{3}{*}{BrainWeb7} & Patch ($k=1$)               & 0.7065                   & 28.39                      \\
                         & Seg. (SAM, $k=1$)              & 0.6972                   & 28.44                      \\
                         & Seg. (MedSAM, $k=1$)              & 0.6940                   & 28.62                      \\ \cmidrule(lr){2-4}
                          & SMoE Patch$^{\star \diamond}$                       & \textbf{0.7262}         & \textbf{29.93}                    \\\midrule
\multirow{3}{*}{BrainWeb9} & Patch ($k=1$)               & 0.6406                   & 25.64                      \\
                         & Seg. (SAM, $k=1$)              & 0.6334                   & 25.59                      \\
                         & Seg. (MedSAM, $k=1$)              & 0.6288                   & 25.69 \\ \cmidrule(lr){2-4}
                        & SMoE Patch$^{\diamond}$                 & 0.6388          & \textbf{26.65} \\
                        & SMoE Seg. (SAM) $^{\star}$           & \textbf{0.6422}          & 25.92      
                        \\ \bottomrule
\end{tabular}%
}
\end{table}

Lastly, we examine \textbf{how much of the performance improvement of our method is due to image decomposition and how much is due to the mixture-of-experts}.
Could a single expert trained on smaller patches and then reconstructed (\(k=1\)) still offer performance improvements over baselines such as DnCNN?
Decomposition alone allows the model to focus on a single spatial region at a time, and allows multiple opportunities to correctly estimate the noise in any overlapping regions.
To explore this, we trained models using all three decomposition methods with a single expert ($k=1$ models), naming them Patch ($k=1$), Seg. (SAM, $k=1$), and Seg. (MedSAM, $k=1$). The results are summarized in Table~\ref{k1brainweb_results} and Table~\ref{k1ixi_results}. 
In most cases, the mixture-of-experts (MoE) models outperform their single-expert counterparts in terms of PSNR. Notably, single-expert models perform exceptionally well when trained on BrainWeb and tested on IXI. However, in terms of SSIM, the performance of single-expert models is generally comparable to that of MoE models.

These results raise an important question: why do decomposition models with a single expert sometimes perform as well as, or even rival, multi-expert models? 
We hypothesize that MRI images exhibit a limited number of inherent noise clusters, allowing a single DnCNN to approximate noise distributions by internally modeling a MoE behaviors to some extent. 
Since $k=1$ models use a single expert to train on all patches, the denoised patches tend to have a more consistent structural appearance compared to those processed by multiple experts, resulting in comparable SSIM values, and sometimes even higher. 
However, this consistency comes at the cost of granularity, as $k=1$ models rely on just one expert and are less capable of adapting to variations in different noise distributions. As a result, they suffer a decrease in PSNR in most cases.
In contrast, the proposed MoE approach explicitly models region-specific noise distributions, providing both granularity and adaptability.

\begin{table}[t]
\centering
\caption{Single-expert models trained over \textbf{IXI}. For comparison, $\star$ denotes the SMoE model with the best SSIM, while $\diamond$ denotes the SMoE model with the best PSNR.}
\label{k1ixi_results}
\resizebox{\columnwidth}{!}{%
\begin{tabular}{@{}lcccccc@{}}
\toprule
\textbf{Test Data}       & \textbf{Model}        & \textbf{SSIM}    & \textbf{PSNR (dB)} \\ \midrule
\multirow{3}{*}{IXI}     & Patch ($k=1$)               & 0.9164                   & 44.53                      \\
                         & Seg. (SAM, $k=1$)              & 0.9337                   & 45.56                      \\
                         & Seg. (MedSAM, $k=1$)              & 0.9194                   & 45.04                      \\ \cmidrule(lr){2-4}
                         & SMoE Patch$^{\star \diamond}$                      & \textbf{0.9383} & \textbf{46.60}                      \\\midrule
\multirow{3}{*}{BrainWeb} & Patch ($k=1$)               & 0.8798                   & 34.93                      \\
                         & Seg. (SAM, $k=1$)              & 0.8950                   & 36.50                      \\
                         & Seg. (MedSAM, $k=1$)              & 0.8937                   & 36.15 \\ \cmidrule(lr){2-4}
                        & SMoE Patch$^{\star}$                & \textbf{0.9025} & 37.19 \\ 
                         & SMoE Segment (SAM)$^{\diamond}$   & 0.8889         & \textbf{37.81}  
                         \\ \bottomrule
\end{tabular}%
}
\end{table}

\section{Conclusion and Future Work}

In this work, we have proposed and evaluated a sparse mixture-of-experts approach to non-uniform noise reduction in MRI images.
Our approach decomposes an image into fine-grained regions and denoises each region with an appropriate expert denoising model.
Our analysis indicates that the higher granularity achieved with our approach out-performs baselines which use coarser decompositions.

We conjecture that our proposed approach is applicable to other types of medical images which also exhibit non-uniform noise, e.g. CT and X-ray imaging, but we leave this to future work.
Non-medical vision applications which are subject to non-uniform corruptions or noise patterns may also benefit from our approach, such as vision sensors placed in autonomous vehicles which experience localized distortions, e.g. glare, raindrops, and dirt.
Additionally, the complexity of the DnCNN architecture may be excessive for cluster-specific denoising.
Research on model pruning or lightweight designs could help maintain performance while reducing computational costs, making these methods more suitable for real-time applications.


{\small
\bibliographystyle{ieee_fullname}
\bibliography{egbib}
}

\end{document}